\documentclass[aps,prl,twocolumn,superscriptaddress]{revtex4}

\usepackage{graphicx}

\begin{document}


\title{The mid-infrared Hall effect in optimally-doped
Bi$_2$Sr$_2$CaCu$_2$O$_{8+\delta}$}

\author{D. C. Schmadel}
\email[]{schmadel@physics.umd.edu}
\affiliation{Department of Physics, University of Maryland, College Park,
Maryland 20742 USA}

\author{J. J. Tu}
\affiliation{Department of Physics, City University of New York, New York, New
York 11973 USA}
\affiliation{Physics Department, Brookhaven National Laboratory, Upton, New
York 11973 USA}



\author{G. D. Gu}
\affiliation{Department of Physics, Brookhaven National Laboratory, Upton, New
York 11973 USA}

\author{Hiroshi Kontani}
\affiliation{Department of Physics, Nagoya University, Furo-cho, Nagoya 464-8602, Japan}

\author{H. D. Drew}
\affiliation{Department of Physics, University of Maryland, College Park,
Maryland 20742 USA}
\affiliation{Center of Superconductivity Research, University of Maryland,
College Park, Maryland 20742, USA}


\date{\today}

\begin{abstract}
Heterodyne polarometry is used to measure the frequency dependence in the mid IR
from 900 to 1100 cm$^-1$ and temperature dependence from 35 to 330 K of the normal state Hall transport in single crystal, optimally doped Bi$_2$Sr$_2$CaCu$_2$O$_{8+\delta}$. The results show a simple Drude behavior in the Hall conductivity $\sigma_{xy}$ which stands in contrast to the more complex, extended Drude behavior for the longitudinal conductivity $\sigma_{xx}$. The mid IR Hall scattering rate $\gamma_{xy}$ increases linearly with temperature and has a small, positive, projected intercept at $T=0$. The longitudinal scatter rate, in contrast, is much larger and exhibits very little temperature dependnece. The Hall frequency indicates a carrier mass which is 6.7 times the band mass and which decrease slighly with increasing frequency. These disparate behaviors are consistent with calculations based on the fluctuation-exchange interaction using current vertex corrections..
\end{abstract}

\maketitle

The experimental system of the current work measures the very small complex
Faraday angle imparted to CO$_2$ laser radiation traveling perpendicular to and
transmitted by the sample immersed in a perpendicular magnetic field.  The system is the same as that used by in the earlier YBCO study~\cite{CernePRL2000} with the addition of an inline calibration system and a continuous stress-free temperature scan provision~\cite{Cerne2002}. The sample of the current work was cleaved or, rather, peeled from a bulk single
crystal of Bi$_2$Sr$_2$CaCu$_2$O$_{8+\delta}$ grown by the travelling
floating zone method.  The resulting 100 nm thick film was placed in thermal contact with a supporting wedged BaF$_2$ crystal.  Measurement of the AC magnetic susceptance of this mounted, peeled segment revealed a T$_{\text{c}}$ of 92 K with a width of less than 1K.  This measurement was performed after all of the Hall measurements had been completed thus establishing the integrity of the sample and recommending the Hall data as representative of optimally doped BSCCO.  Infrared conductivity data~\cite{TuPRB2002} from measurements performed on bulk crystals from the same batch supplied the real and imaginary parts of $\sigma_{xx}$ required to obtain the Hall angle from the Faraday angle data.  The calculation of $\sigma_{xy}$ considered the multilayer reflection effects within the BSCCO film~\cite{Cerne2002}.

\begin{figure}
\includegraphics[width=8.6cm, clip=true]{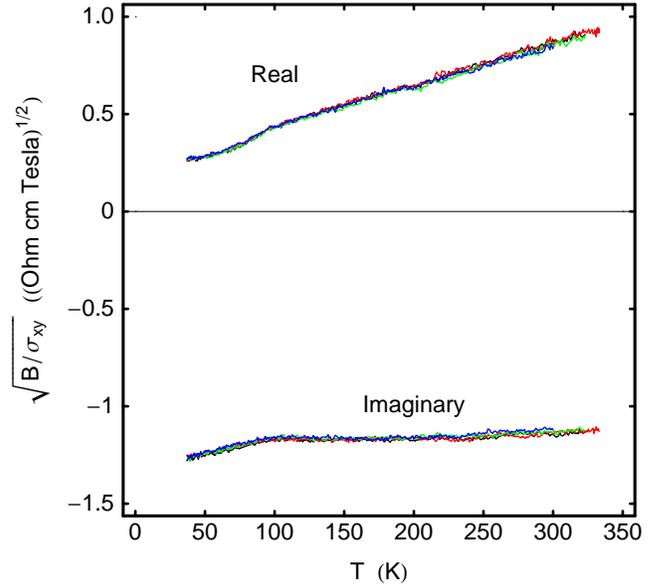}
\caption{\label{fig;dataSigma} The real and imaginary parts of
$1/\sqrt{\sigma_{\text{xy}}}$ for 2212 BSCCO versus temperature and normalized to one Tesla.}
\end{figure}

\begin{figure}
\includegraphics[width=8.6cm, clip=true]{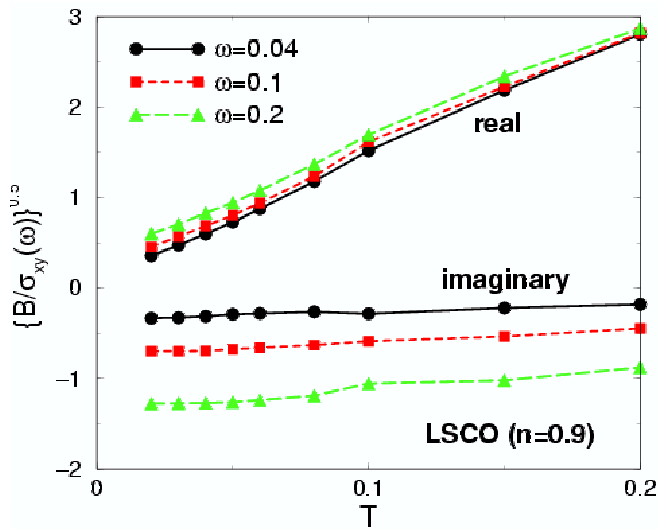}
\caption{\label{fig;L-invsxy-T}$\text{Im}\left(\sigma_{xx}\right)/\text{Re}\left(\sigma_{xx}\right)$
vs. frequency for single crystal 2212 BSCCO at different temperatures.}
\end{figure}

In a simple Drude model the Hall conductivity can be represented by 

\begin{equation}
\sigma_{\text{xy}}=\frac{\omega_p^2}{4\pi}\frac{\omega_H}{(\gamma-i\omega)^2}.
\end{equation}
The real and imaginary parts of the inverse square root of this Drude Hall conductivity relate directly to the Hall scattering rate and mass. For this reason we plot $1/\sqrt{\sigma_{\text{xy}}}$ from the data in  figure~\ref{fig;dataSigma}. The most striking feature is the strong temperature dependence of the real part in comparison with that of the imaginary part. Remarkably, this behavior is found the calculated results of the FLEX+CVC ~\cite{KontaniFLEX+CVC} approximation shown in Fig.~\ref{fig;L-invsxy-T}. This calculation involved the fluctuation-exchange FLEX approximation using current vertex corrections CVC, which satisfy the conservation laws.

\begin{figure}
\includegraphics[width=8.6cm, clip=true]{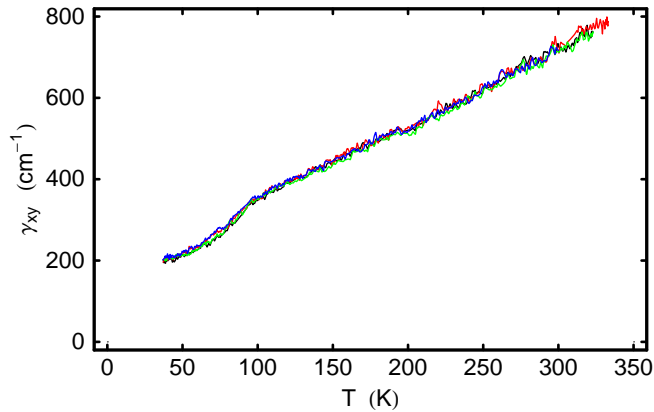}
\caption{\label{fig;dataGamma} The Hall scattering rate 
$\gamma_{xy}$ versus temperature for 2212 BSCCO.}
\end{figure}

Using this model, we calculated the Hall scattering rates and Hall mass from the data. The results are shown in Fig.~\ref{fig;dataGamma}. The Hall scattering rate varies linearly with temperature from 400 to 800 $cm^{-1}$. The scattering rate calculated from the longitundinal conductivity~\cite{TuPRB2002} is considerably greater and exhibits very little temperature dependence over this range--rising from 1400 $cm^{-1}$ at low temperature to 1600 $cm^{-1}$ at room temperature. The Hall scattering rate also has a rather small projected intercept of only 200 $cm^{-1}$ at zero degrees. For the high frequency of the mid IR, this indicates a very small frequency contribution to the scattering rate, unlike the longitudinal scattering rate, which rises linearly with frequency.

The Hall mass calculated using the simple Drude model $eb/mc$ and a plasma frequency of 17,000 $cm^{-1}$ is 6 $m_e$ as compared with 2 $m_e$ calculated from the longitundinal conductivity~\cite{TuPRB2002}. Further, the Hall mass decreases slowly with temperature. (We are currently cheeking the frequency dependence.)

This work was supported by the NSF under grant DMR-0030112

\bibliography{BSCCOMidHall102705}

\end{document}